\documentclass[prl,showpacs,footinbib,twocolumn,final]{revtex4}
\usepackage{color,amstext,amsxtra,amssymb,latexsym,epsfig}

\newcommand{\K}{{\boldsymbol K}}

\renewcommand{\a}{{\boldsymbol a}}
\renewcommand{\k}{{\boldsymbol k}}

\newcommand{\be}{\begin{equation}}
\newcommand{\ee}{\end{equation}}

\newcommand{\p}{{\boldsymbol p}}

\newcommand{\D}{{\boldsymbol D}}

\newcommand{\R}{{\boldsymbol R}}

\newcommand{\ep}{\epsilon}

\definecolor{gris}{gray}{0.9}

\begin{document}
\title{A new magnetic field dependence of Landau levels 
\\ on a graphene like structure}


\author{Petra Dietl, Fr\'ed\'eric Pi\'echon and Gilles Montambaux}
\affiliation{Laboratoire de Physique des Solides, Universit\'{e}
Paris Sud, CNRS UMR8502, 91405 Orsay Cedex, France}

\date{\today}


\begin{abstract}
We consider a tight-binding model on the honeycomb lattice in a
magnetic field. For special values of the hopping integrals, the
dispersion relation is linear in one direction and quadratic in
the other. We find that, in this case, the energy of the Landau
levels varies with the field $B$ as $\epsilon_n(B) \sim
[(n+\gamma)B]^{2/3}$. This result is obtained from the low-field
study of the tight-binding spectrum on the honeycomb lattice in a
magnetic field (Hofstadter spectrum) as well as from a calculation
in the continuum approximation at low field. The latter  links the new spectrum to the one of a modified quartic
oscillator. The obtained value $\gamma=1/2$ is found to result
from the cancellation of a Berry phase.
\end{abstract}

\maketitle

{\it Introduction - } The recent discovery of graphene has boosted
the study of the physical properties of the honeycomb lattice,
especially in a magnetic field \cite{Novoselov1}. Among the
peculiarities  of the electronic dispersion relation, the spectrum
in the band center $(\ep=0$) is linear, exhibiting the
so-called Dirac spectrum around two  special points at the corners $\K$ and $\K'$ of the Brillouin zone. Near these points the density
of states varies linearly \cite{Wallace}. In a magnetic field, it
has been shown \cite{McClure}  that the energy levels around
$\ep=0$ vary  with the field $B$ as $\epsilon_n(B) \sim \pm [n
B]^{1/2}$, with a degeneracy which is twice the degeneracy of
usual Landau levels (LLs). This is easily explained by the
two-fold valley degeneracy corresponding to the two points $\K$ and
$\K'$. This spectrum has to be contrasted with the familiar field dependence
of Landau levels $\epsilon_n(B)=(n+1/2)e B/m$ for electrons in a
quadratic band with mass $m$. This square root dependence has been
observed experimentally
\cite{Sadowski,QHE}.

Here we present an example where the field dependence of the LLs is
neither linear nor a square root, but reveals a new power-law,
namely a $[(n+\gamma)B]^{2/3}$ behavior, with $\gamma=1/2$. This
is obtained for tight-binding  electrons on the honeycomb lattice,
the same problem as for graphene, but with  special values of the
hopping integrals between nearest neighbors which, contrarily to
the case of graphene, are not taken to be equal. We find that
around a special point of the reciprocal space, the zero-field
spectrum is linear in one direction and quadratic in the other.
This "hybrid" spectrum leads to a new magnetic field dependence of
the Landau levels, properly described by a quartic oscillator
$V(X)=X^4+2X$. The value $\gamma=1/2$ is found to result from the
cancellation of a Berry phase.
\medskip

\begin{figure}[ht]
\begin{center}
{\epsfxsize 4.5cm \epsffile{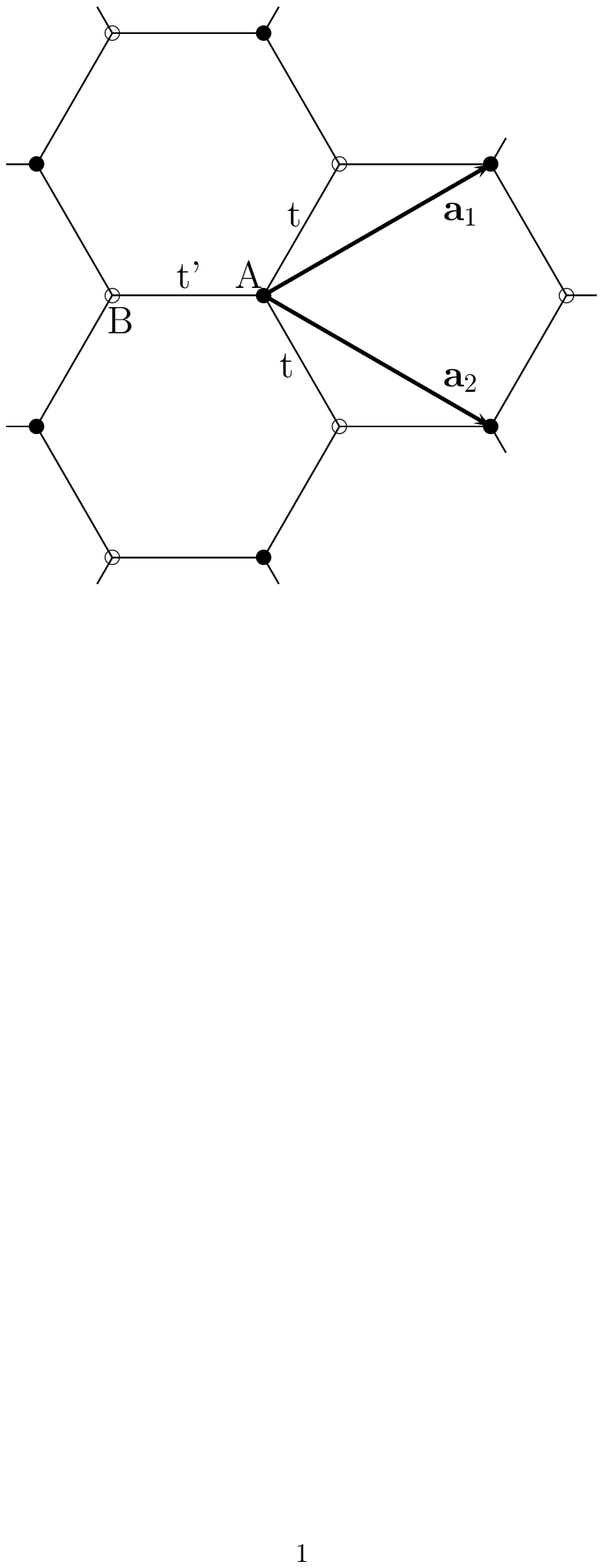}}
\end{center}
\vspace{-1cm}
\parbox[t]{4cm} { \hfill {\it a)}}
\vspace{-.5cm}
\begin{center}
{\hfill \parbox[t]{4cm} {\begin{center}{\epsfysize 3cm
\epsffile{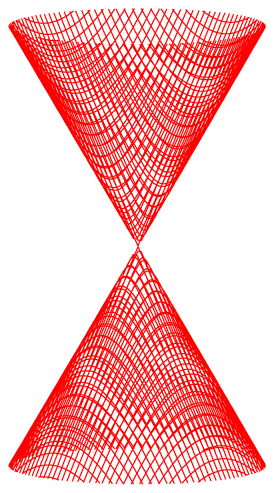}}
\end{center}}
 \hfill
\parbox[t]{4cm}{\begin{center}
{\epsfysize 3cm
\epsffile{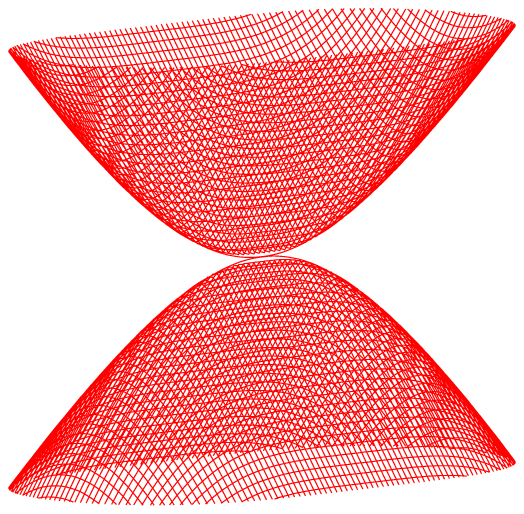}}\end{center}}   \hfill }
\end{center}
\vspace{-1cm}
\parbox[t]{4cm} { \hfill {\it b)}} \parbox[t]{4cm} { \hfill {\it c)}}
 \caption{\em a) Honeycomb lattice with hopping integrals $t$ and $t'$, and elementary
 vectors $\a_1$ and $\a_2$ discussed in the text. b,c)
 Low energy spectrum for the cases $t'=t$ (b) and $t'=2 t$ (c).}
 \label{reseau}
\end{figure}

{\it The model - } We consider the tight-binding  model on the
honeycomb lattice with the possibility that one of the three
hopping elements between nearest neighbors may take a different
value $t'$ from the two others $t$ (fig. \ref{reseau}). This
problem has been studied recently both in zero field
\cite{Hasegawa0} and in a magnetic field \cite{HasegawaH}, where
the authors mention the special interest of the case $t'=2t$. For
these parameters, the authors mention a square-root energy
dependence of the density of states at the  band center
\cite{Hasegawa0}, and they compute the evolution of the Hofstadter
spectrum when $t'$ varies between $t$ and $2t$ \cite{HasegawaH}.
Here, we emphasize the new and peculiar character of the LLs around
$\ep=0$, which has not been foreseen in previous works.

The tight-binding Hamiltonian couples sites of different
sublattices named $A$ and $B$. The eigenvectors are Bloch waves of
the form
\be |\k \rangle={1 \over \sqrt{N}} \sum_j \left(\, c_\k^A |\R_j^A
\rangle +c_\k^B |\R_j^B \rangle \, \right) e^{i \k.\R_j}
\label{bloch} \ee
where $|\R_j^A\rangle, |\R_j^B\rangle$ are atomic states. The sum
runs over vectors of the Bravais lattice. The eigenequations read
\begin{eqnarray}
\epsilon c_\k^A &=& - [t e^{i \k \cdot \a_1} + t e^{i \k \cdot
\a_2} + t' ] c_\k^B \label{eigeneqs} \\
\epsilon c_\k^B &=& - [t e^{-i \k \cdot \a_1} + t e^{-i \k \cdot
\a_2} + t' ] c_\k^A \nonumber
\end{eqnarray}
where $\a_1=a({3  \over 2}, {\sqrt{3} \over 2})$, $\a_2=a({3 
\over 2},-  {\sqrt{3} \over 2})$  are elementary  vectors of the
Bravais lattice, $a$ is the interatomic distance, and $t$, $t'$
are shown in fig. \ref{reseau}.a.  It is well known that when
$t'=t$, the energy vanishes at two points $\D$ and $\D'$ which are located at the corners $\K$ and $\K'$ of the Brillouin zone ($\K= 2 \a_1^*/3 + \a_2^*/3$, $\K'=
\a_1^*/3 + 2 \a_2^*/3$, where $\a_1^*$ and $\a_2^*$ are reciprocal
lattice vectors). As $t'$ increases, the two points $\D$ and $\D'$
approach each other (their distance varies as $\overrightarrow{DD'}={3
\over \pi} \overrightarrow{KK'}  \arctan \sqrt{ 4 t^2/t'^2 -1}$) and
merge into a single point $\D_0=(\a_1^*+\a_2^*)/2$ when $t'=2t$ (for
$t'
> 2t$, a gap opens between the two subbands).
 An expansion
$\k=\D+\p$ around these two points gives the low-energy spectrum,
solution of the two equations (here $a=1$  for shorter notations)

$$ \epsilon c_\k^A= \big[-{3 \over 2} i t' p_x   \pm
{\sqrt{3}} t'' p_y  + {3\over 8} t'(3 p_ x^2 + p_y^2)+ {3 \sqrt{3} \over 2} t'' p_x p_y \big] c_\k^B $$
 with $t''=\sqrt{ t^2 - {t'^2 / 4}}$ and a similar expression for
$\epsilon c_\k^B$ with $(i \rightarrow - i)$. The  $\pm$ sign denotes
the vicinity of the two points $\D$ and $\D'$.

We now consider the case $t'=2 t$. In  the vicinity of the single
point $\D_0$, keeping the leading terms in each direction, we find
\begin{equation}
 \epsilon c_\k^A = t a ( -3 i p_x + {3 \over 4} a p_y^2) c_\k^B
\end{equation}
and a similar equation with $(A \rightarrow B, i \rightarrow - i
)$, so that the Hamiltonian can be written in the form
\be H=\left(%
 \begin{array}{cc}
  0 &  - i c p_x + { p_y^2 \over 2 m} \\
   i c p_x +  {p_y^2 \over 2 m }& 0 \\
\end{array}%
\right) \label{hamilt} \ee
where we have defined the velocity $c= 3 t a$, an effective mass
$m= 2/(3 t a^2 )$, and a "mass energy" $m c^2 = 6 t$ (we fix
$\hbar=1$). The eigenvalues read

\be \epsilon = \pm  (c^2 p_x^2 +    {p_y^4\over 4 m^2})^{1/2} \ .
\label{epxpy} \ee

Remarkably, the spectrum is linear in one direction, quadratic in
the other (fig. \ref{reseau}.b,c). It has been noticed
\cite{Hasegawa0} that such a dispersion relation leads to a square
root dependence of the  density of states, that can be written  in
the form $\rho(\epsilon)\propto {\sqrt{ m } \over c} \sqrt{\epsilon} $
which is unusual in two dimensions. One may expect that this
peculiar behavior leads to a new repartition of LLs.\medskip

{\it Effect of the magnetic field, new Landau levels - }
To describe qualitatively the effect of a weak  magnetic field, we
first start with a simple semiclassical argument. The quantization
condition for energy levels in a magnetic field $B$ has the form
$S(\ep)=2 \pi (n+\gamma) e B$ where $S(\ep)$ is the area of an
orbit of energy $\ep$ in reciprocal space and $\gamma$ is a
constant $0 \leq \gamma < 1$ \cite{Onsager}. From eq. (\ref{epxpy}),
one finds easily \cite{remdos}

 \be S(\ep) = \beta {\sqrt{m} \over c} \ep^{3/2}\quad  \longrightarrow \quad \ep_n
=\left({2 \pi  e c \over \beta \sqrt{m}}\right)^{2/3} [(n+\gamma)
B ]^{2/3} \label{sclassics} \ee
with $\beta=2 \, \Gamma(1/4)^2/(3 \sqrt{\pi}) \simeq
4.9442$. This new behavior has to be contrasted with the usual
case of free massive particles where $S(\ep)=2 \pi m \ep$, so that
$\ep_n = \omega_c(n+\gamma)$ or with the case of Dirac particles
where $S(\ep)= \pi \ep^2/c^2$, leading to  a
square root magnetic field dependence $\ep_n=\pm c \sqrt{2 e
B(n+\gamma})$ of the energy levels. The phase factor $\gamma$ cannot be obtained from
such semiclassical argument.

The $B^{2/3}$ dependence can also be obtained from another simple
argument related to the new energy dependence of the density of
states. Since it has a square root dependence near $\ep=0$, one may expect a new
field dependence of the LLs, simply because the
repartition of the zero field states into LLs is
different. Assume quite generally a power-law dependence of
the zero field density of states, $\rho(\epsilon) \sim
\epsilon^\kappa$ and a given number $n$ of filled LLs.
The total number of states in these levels, $n e B/h$, has to be
equal to the total number of filled states which is or order
$\int_0^{\epsilon_n(B)} \rho(\ep) d \ep$ so that
$\epsilon_n(B) \sim [(n+\gamma) B]^{1\over \kappa +1}$ \cite{Moessner}. A  constant density
of states leads to the usual LLs, a linear density of
states to the square-root behavior observed in graphene.  Here, the same  
argument gives  a new power law $\epsilon_n(B) \sim [(n+\gamma) B]^{2\over
3}$ for the LLs.

\begin{figure}[ht]
\begin{center}
{\hfill \parbox[t]{4cm}{\epsfxsize \hsize \epsffile{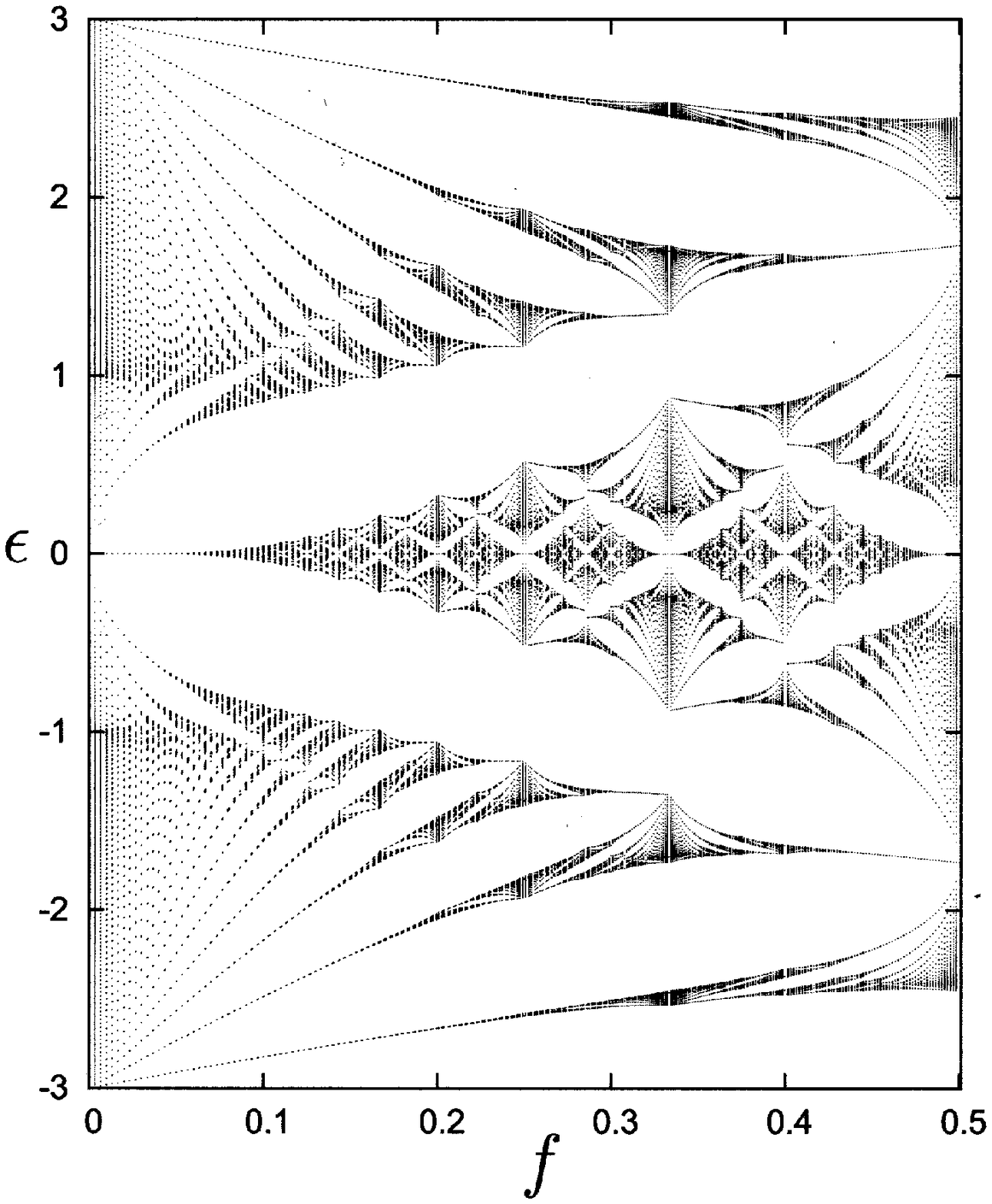}}
\hfill
\parbox[t]{4cm}{\epsfxsize \hsize
\epsffile{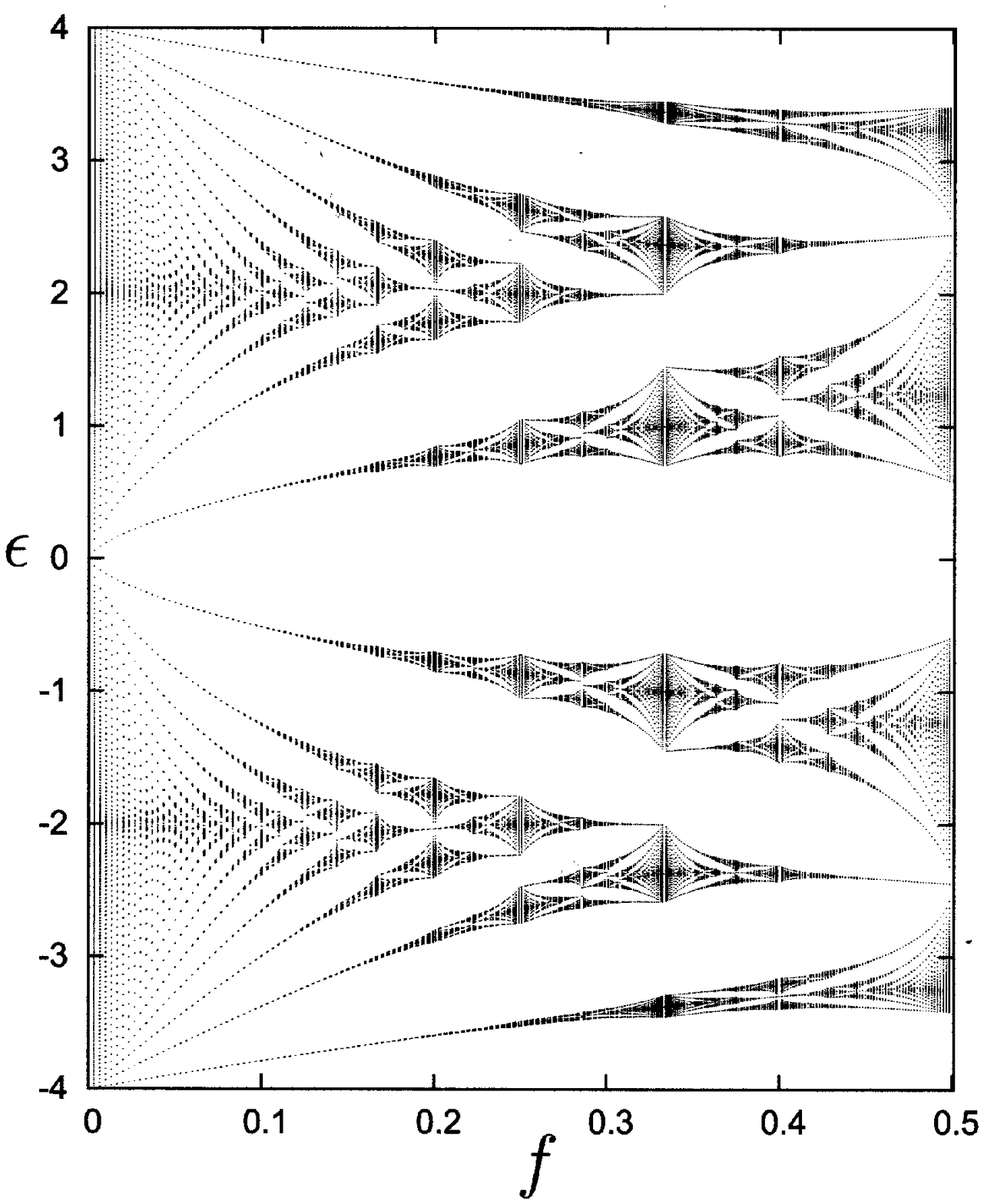}}   \hfill } \end{center}
\vspace{-.5cm}
\parbox[t]{4cm} { \hfill {\it a)}} \parbox[t]{4cm} { \hfill {\it b)}}
\begin{center}
{\hfill \parbox[t]{4cm}{\epsfxsize \hsize
\epsffile{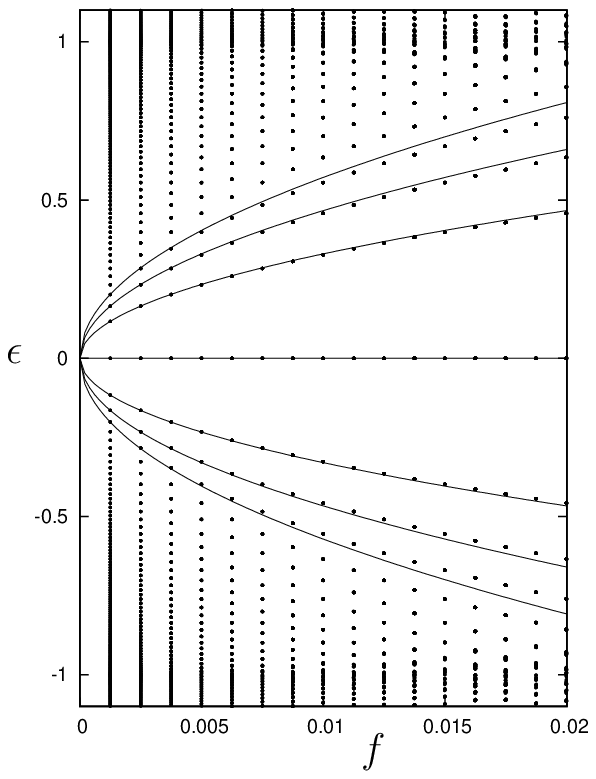}} \hfill
\parbox[t]{4cm}{\epsfxsize \hsize
\epsffile{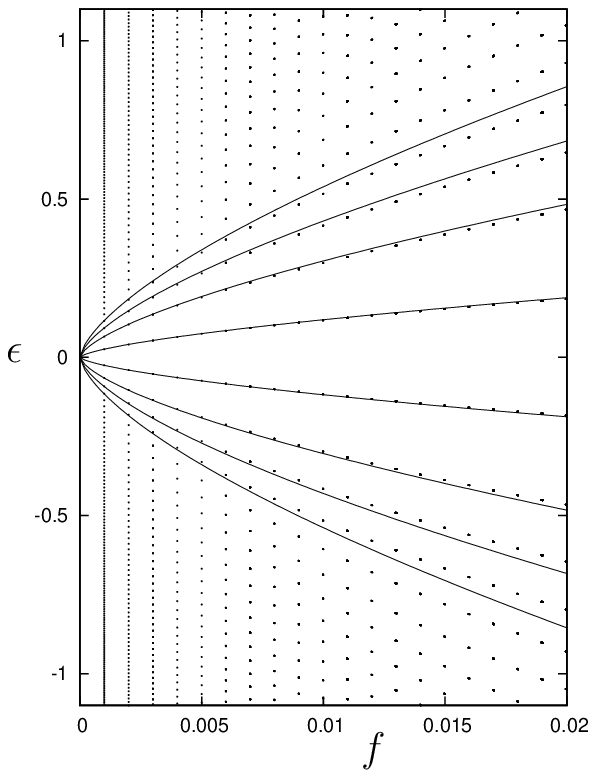}}   \hfill } \end{center}
\vspace{-.5cm}
\parbox[t]{4cm} { \hfill {\it c)}} \parbox[t]{4cm} { \hfill {\it d)}}
\caption{\em a)  Hofstadter-Rammal spectrum in the case $t'=t$. b)
Spectrum for the case $t'=2t$. c)  Low-field behavior in the case
$t'=t$. The low-field bands are well fitted by the analytical
expansion $\ep_n=\pm t \sqrt{2 \pi 3^{1/2} n f }$. d) In the case
$t'=2t$, the low-field spectrum is well fitted by eq.
(\ref{spectrexact}). Deviations at higher field or higher energy
are due to lattice effects.}
 \label{spectrum}
\end{figure}

Fig. \ref{spectrum} presents the spectrum of the tight-binding
problem in a magnetic field (the so-called Hofstadter spectrum)
for the honeycomb lattice. It has first been calculated by R.
Rammal for the case $t'=t$ and more recently for $t'=2t$
\cite{Rammal,HasegawaH}. The procedure to obtain this spectrum is
described at length in ref. \cite{Rammal}. The fractal structure
results from the competition between magnetic field and lattice
effects. For a commensurate reduced flux $f=p/q$, the spectrum
exhibits $2q$ sub-bands. It is a periodic function of the reduced
flux $f= B a^2 3\sqrt{3}/(2 \phi_0)$  through one plaquette in
units of the flux quantum $\phi_0$.  At low field, the lattice
effects are negligible, and we expect to recover the results of a
continuum limit. The linear dependence of the LLs is
clearly seen on the top and bottom of the spectrum (fig.
\ref{spectrum}). For $t'=t$, the square-root dependence of the
levels is observed around $\ep=0$. For $t'=2t$, the levels exhibit
the new magnetic field dependence which is well fitted by a power
law $B^{2/3}$ (fig. \ref{spectrum}.d).
In order to derive the low-field spectrum analytically, we now calculate the magnetic field effect in the continuum
approximation around $\ep=0$.
\medskip

\begin{figure}[ht]
\centerline{ \epsfysize 5cm \epsffile{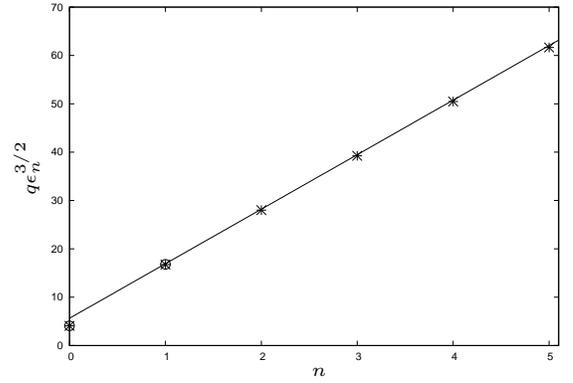} } 
\caption{ For
$f=1/q$, plot of $ q \epsilon_n^{3/2}$ vs $n$ ($*$), and
comparison with the WKB solution (\ref{WKB2}) of the quartic
oscillator (straight line) and with the numerical solution of the
modified quartic, $X^4+2X$ oscillator (o) for $n=0,1$. For
 $n \geq 2$, the two solutions are indistinguishable.} 
\label{fit}
\end{figure}

{\it Spectrum in the continuum approximation - } We now apply a
magnetic field $B$ and use the Landau jauge $\boldsymbol{A} = (0,B
x, 0)$. The substitution $p_y \rightarrow p_y - e B x$
leads to the new Hamiltonian

$$H=\left(%
 \begin{array}{cc}
  0 &  - i c p_x + {1 \over 2} m \omega_c^2 \tilde x^2 \\
   i c p_x + {1 \over 2} m \omega_c^2 \tilde x^2 & 0 \\
\end{array}%
\right)
$$with an effective cyclotron frequency $\omega_c= e B /m=3 e B t a^2/2$
which can be also written in terms of the reduced flux $f= B a^2
3\sqrt{3}/(2 \phi_0)$  through one plaquette of the lattice,
namely $ \omega_c ={2 \pi \over \sqrt{3}} t f$. In $\tilde x= x
-p_y/eB$, quantization of $p_y$ leads to the usual degeneracy of
Landau levels.
 The energy
levels $\epsilon$ are solutions of

$$\Big( c^2 p_x^2 + ({1 \over 2} m \omega_c^2)^2 \tilde x^4 -i {c \over 2} m \omega_c^2 [p_x,\tilde x^2]\Big) \psi=\epsilon^2 \psi \ . $$
Introducing new dimensionless conjugate variables $X$ and $P$, we
rewrite
\begin{equation}
\left({m \omega_c^2 c^2 \over 2 }\right)^{2/3} \Big( P^2 + X^4 - i
[P, X^2]\Big) \psi=\ep^2 \psi \ . \label{qosc} \end{equation}
 This expression shows that the
eigenvalues necessarily scale as $B^{2/3}$.    Neglecting first the linear term $[P,
X^2]= - 2 i X$,  the
eigenvalues of the quartic oscillator $P^2 + X^4$ can be  easily 
estimated, at least for large $n$, in the WKB approximation
\cite{Montroll} and are found to be of the form $C(n+1/2)^{4/3}$
with $C=[3 \pi \sqrt{2 \pi} /\Gamma(1/4)^2]^{4/3} \simeq
2.18507$. Therefore,  the Landau levels are given by
(restoring $\hbar$)

\be \epsilon_n=\pm A
 (m c^2)^{1/3} [(n+1/2)\hbar \omega_c]^{2/3} \label{WKB1} \ee
with $A=\sqrt{C \over 2^{2/3}}\simeq 1.17325$. This is precisely
the dependence expected from the above semiclassical argument
(\ref{sclassics}), with the phase factor $\gamma$ now determined
to be $1/2$. Replacing $m$ and $\omega_c$ by their expressions in
terms of the lattice parameters considered here, we finally obtain 
\be \epsilon_n = \pm \alpha [(n+1/2) f]^{2/3} \label{WKB2} \ee
where $\alpha= (2 \pi)^{2/3} \sqrt{C} \simeq 5. 0333$. The
$f^{2/3}$ dependence of these levels is clearly seen in fig.
\ref{spectrum}.d, and their $(n+1/2)^{2/3}$ dependence is confirmed on fig. \ref{fit}.

It is interesting to  compare eq. (\ref{qosc}) with the
eigenequations for Dirac fermions:

\be e B c^2 \Big( P^2 + X^2 - i [P,X]\Big) \psi=\epsilon^2 \psi  \ , 
\label{hosc} \ee
 with $\epsilon_n=\pm c \sqrt{2 e B n}$ or for
free massive particles
\be{  \omega_c \over 2}  \left(P^2+X^2
\right) \psi=\epsilon \psi \ , \label{hosc0} \ee
with $\epsilon_n=(n+1/2) \hbar \omega_c$, and to emphasize the
role of the commutator which enters eqs (\ref{qosc},\ref{hosc}).
There are several discussions in the literature
\cite{Mikitik} to explain the disappearance of the phase term
$1/2$ in the case of the Dirac spectrum, and we return to this
point later. We notice from  eq. (\ref{hosc}) that this disappearance
comes simply from the commutator $[P,X]= -i $. In our case, the
commutator which appears in eq. (\ref{qosc}), $[P,X^2]=-2 i X$,
modifies the quartic potential which becomes $X^4 + 2 X$. This
linear term is actually a small perturbation negligible when $n$
is large. Taking into account this linear term, a numerical
calculation of the eigenvalues of this modified quartic oscillator
finally gives (fig. \ref{fit})
\be \ep_n= \pm \alpha g(n) [(n+1/2) f]^{2/3}\label{spectrexact}
\ee
 where $g(0) \simeq 0.808, g(1) \simeq 0.994$. For $n \geq 2$, $g(n) \sim 1$ so that the WKB solution (\ref{WKB2}) of the quartic oscillator turns out to be extremely good. Finally, let us mention that since the two
Dirac points $\D$ and $\D'$ have merged into a single point $\D_0$
for $t'=2t$, the valley degeneracy has disappeared and the LLs degeneracy
has recovered its usual value.

\begin{figure}[ht]
\begin{center}
{\epsfxsize 4cm \epsffile{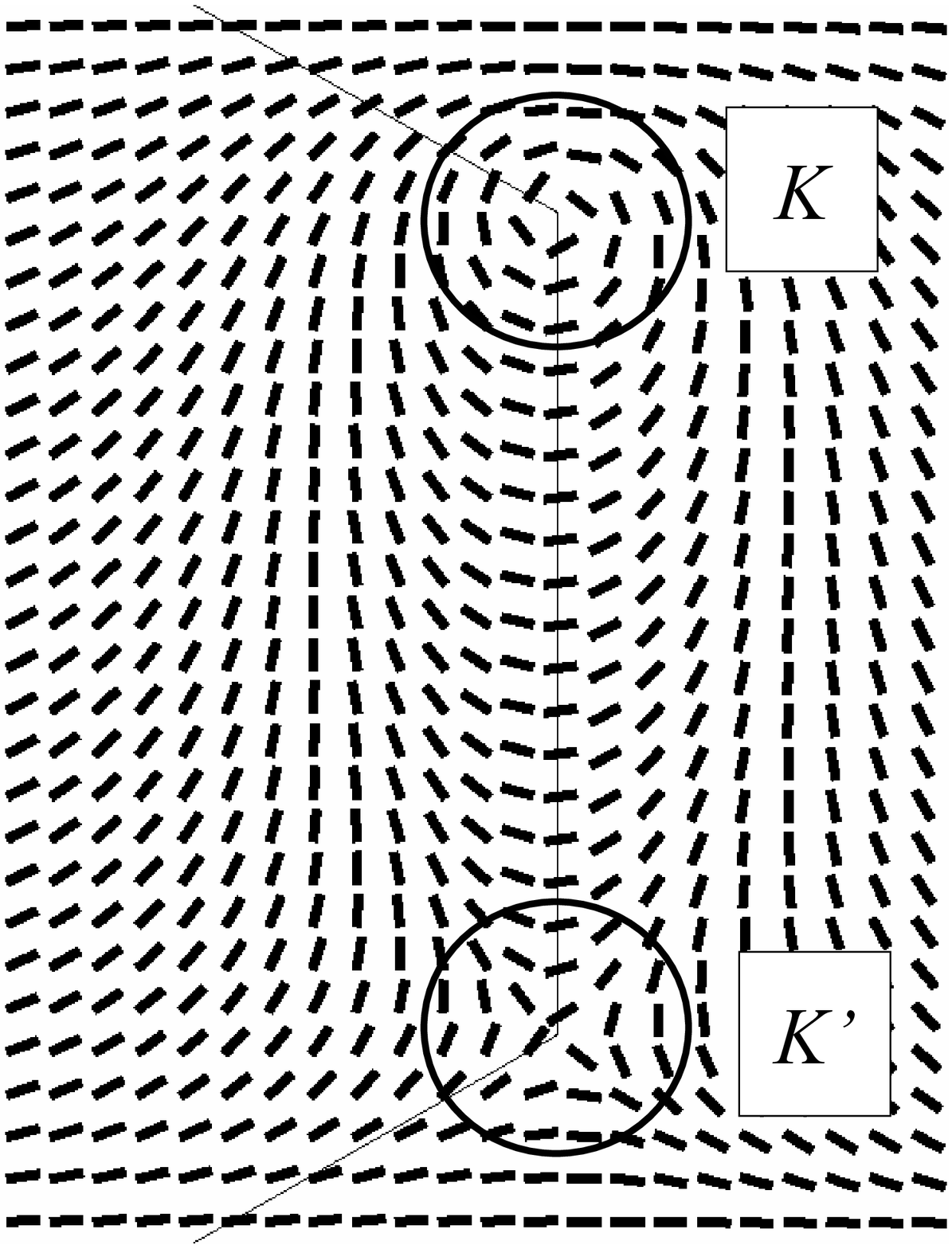} } \hfill {\epsfxsize
4cm \epsffile{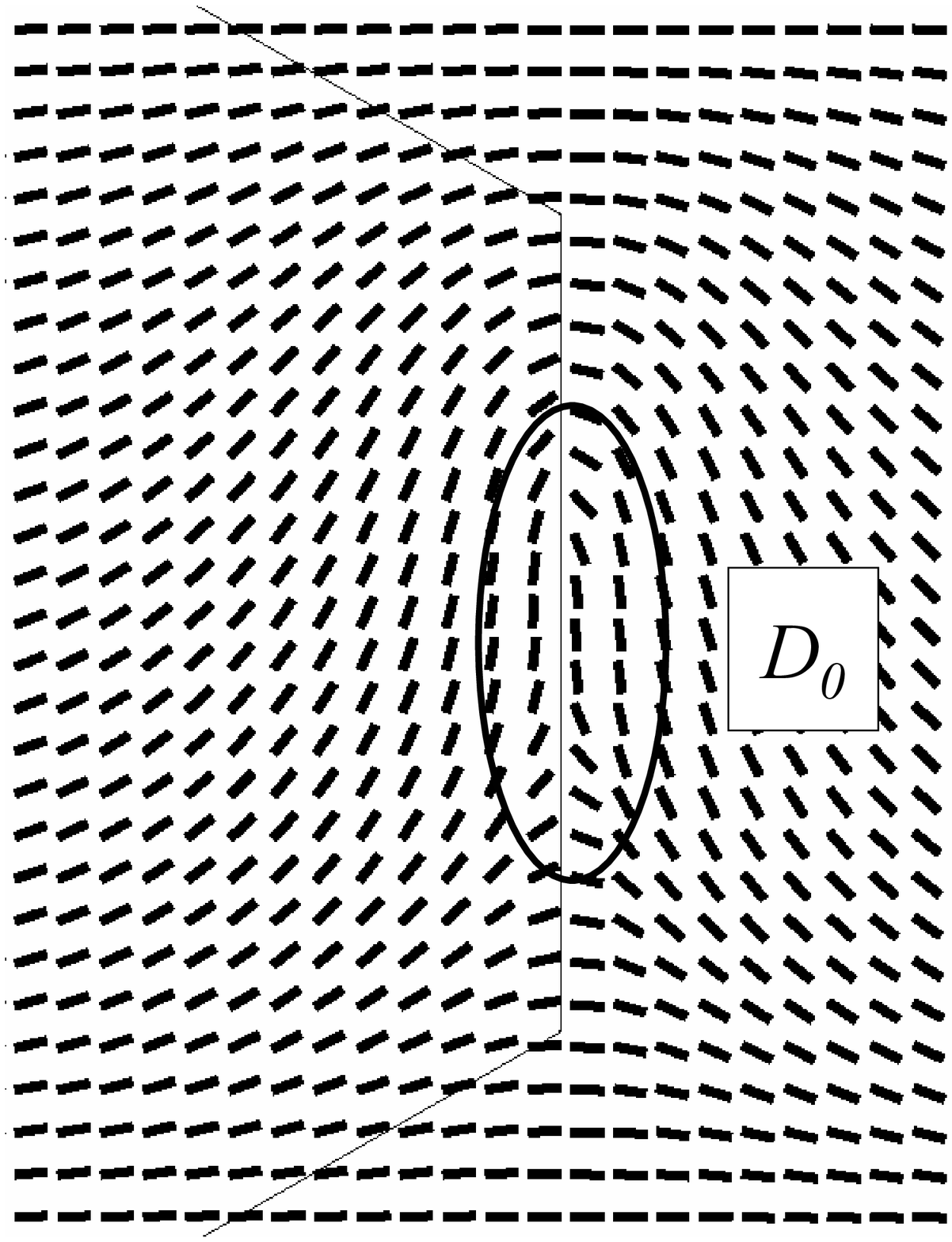}}
\end{center}
\vspace{-.5cm}
\parbox[t]{4cm} { \hfill {\it a)}} \parbox[t]{4cm} { \hfill {\it b)}}
 \caption{\em  Wave-vector dependence of the ratio $c_\k^B/c_\k^A$ for the eigenvectors
 (\ref{bloch}). The figures represent a part of the Brillouin zone near the points $\K$ and $\K'$. a) For $t'=t$, the Berry phase $\pm \pi$ on semiclassical orbits  around the points $\D=\K$, $\D'=\K'$ implies
 $\gamma=0$. For $t'=2t$, the Berry phase on a contour around  $\D_0$ vanishes,
 implying $\gamma=1/2$.}
 \label{berryfig}
\end{figure}

We now comment on the relation between the phase factor
$\gamma$ entering the quantization of semiclassical orbits and a Berry phase. It has been established that  \cite{Mikitik}
$$\gamma={1 \over 2} - {1 \over 2 \pi} \oint_\Gamma \langle \k
|\nabla_\k |\k \rangle \cdot d \k $$
 where $\Gamma$ is the contour of a semiclassical orbit. It is easily found from eqs. (\ref{bloch},\ref{eigeneqs}) that, for the isotropic case $t'=t$,
$\oint_\Gamma \langle \k |\nabla_\k |\k \rangle \cdot d \k=\pm
\pi$  around the points $\D$ and $\D'$ which can be seen as topological defects (see fig. \ref{berryfig}.a), leading to
$\gamma=0$ as well known for graphene. For our case $t'=2t$, these two topological defects $\D$ and $\D'$ merge into a single one $\D_0$ and {\it annihilate}. As may be seen on fig. \ref{berryfig}.b, we now obtain $\oint_\Gamma \langle \k |\nabla_\k |\k \rangle \cdot d \k=0$, which explains why $\gamma=1/2$.
\medskip

{\it Conclusion - } The field dependence of the LLs
depends dramatically on the structure of the zero field dispersion
relation. Recent experiments and calculations for graphene have
shown that a linear dispersion relation leads to a square-root
dependence of the LLs with the field. In this paper, we
have presented an example with an usual dispersion relation, which
leads  to a square-root energy dependence of the density of states
and to a new field dependence $[(n+1/2) B]^{2/3}$ of the LLs. More generally, we may consider a dispersion relation of
the form $\epsilon =(p_x^\alpha +p_y^\beta)^\delta$. 
The area of the orbits of a given energy $\ep$ varies as $S(\ep)
\propto \ep^{\alpha + \beta \over \alpha \beta \delta}$, so that
the Onsager quantization rule for energy levels  leads to the
general dependence of the Landau levels with the magnetic field
$$
\epsilon_n(B) \sim [ (n + \gamma) B ]^{{\alpha \beta \delta \over
\alpha + \beta}} \ .$$
This dependence can also be obtained from a counting argument
based on the energy dependence of the density of states which is
easily found to vary as
$ \rho(\ep) \sim \ep^{{1 \over \alpha \delta}+{1 \over \beta
\delta}-1}$.

To conclude, we briefly comment on possible experimental realizations leading to such an electronic spectrum. An anisotropic version of the graphene structure with $t' \neq t$, called quinoid and discussed long ago by L. Pauling {\cite{Pauling}, could be induced by uniaxial stress or bending of a graphene sheet. We also mention recent discussions on the feasibility of such a structure with cold atoms in an optical lattice created by laser beams \cite{Gremaud}.

\medskip

{\it Acknowlegments - } The authors acknowledge useful
discussions with J. Dalibard, J.N. Fuchs and  M.O. Goerbig.


\begin{thebibliography}{99}
\bibitem{remdos} The exact expression  of
the density of states can be easily recovered from $\rho(\ep)=
{\partial S(\ep) / 4 \pi^2 \partial \ep}$.
\end{thebibliography}
\end{document}